# LE TRADING ALGORITHMIQUE

© Victor Lebreton [*] - février 2007

* Consultant pour les marchés financiers, Doctorant au CES – MATISSE, Université Paris1 Panthéon Sorbonne et Ms Spécialisé Finance de marché et Architecture IT à Ecole Centrale d'Electronique. victor.lebreton@malix.univ-paris1.fr, lebreton@ece.fr.

INTRODUCTION

Le trading algorithmique provient de la dématérialisation du traitement des ordres d'achats ou de ventes d'actifs. Depuis 1980 l'informatisation des places boursières offre des possibilités de traitement en temps réel de l'information financière. Cette révolution technologique a permis de développer des procédés et des méthodes d'évaluations mathématiques pour identifier des moments où les transactions retournent des bénéfices. Les recherches actuelles se portent vers des systèmes de transaction autonomes programmés selon certaines périodes et certains algorithmes. En effectuant des opérations plus rapidement ils offrent des possibilités de gain là où le trader ne peut intervenir. Il existe une trentaine d'algorithmes pour assister le trader, les plus connus sont le VWAP, le TWAP, TVOL. Les algorithmes les plus récents offrent des stratégies décisionnelles et font l'objet de nombreuses recherches. Ces avancées dans la modélisation d'automates décisionnels permettent d'envisager un avenir riche pour ces technologies, les acteurs en faisant déjà usage pour plus de 30 % de leurs techniques de trading.

PRINCIPES DE BASE

Ce que l'on appelle aujourd'hui trading algorithmique se compose de deux activités : les opérations de bourse assistées par des algorithmes qui anticipent et favorisent les opportunités de bénéfices (en informant le trader par des graphiques, des alertes et des traitements automatiques), et le trading automatisé qui utilise des automates comme agents autonomes effectuant des transactions selon des algorithmes et des stratégies paramétrées.

En première partie de cette étude je présenterai brièvement l'histoire de ces technologies, à partir des premiers temps des échanges d'informations aux moyens de relais électriques jusqu'à la naissance des réseaux télématiques. Je présenterai ensuite la mise en place des premiers échanges automatisés. Enfin je parlerai de la période actuelle et de l'informatisation massive des acteurs du marché.

La seconde partie présentera un état de l'art des différents algorithmes actuellement sur le marché. On verra alors qu'il existe plusieurs niveaux d'intervention dans le trading algorithmique, soit grâce aux différentes avancées dans le développement des automatistes, soit par ce que les stratégies de trading ont été perfectionnées. Ce sera alors l'occasion d'aborder plus en détail les aspects stratégiques liés à cette discipline.

Puis je présenterai dans une troisième partie les éléments techniques et théoriques nécessaires à la mise en place de ce type de trading. Il s'agit principalement des environnements informatiques spécifiques et des méthodes utilisées en finance quantitative pour suivre et anticiper les marchés au plus près. Nous verrons que ces méthodes font régulièrement appel à la recherche en mathématique appliquée ou à l'intelligence artificielle.

Ce travail de synthèse a pour but de permettre une meilleure compréhension des techniques de trading actuelles en présentant des informations disponibles dans l'univers anglo-américain. Même si ces technologies ne sont pas encore pleinement développées il semble évident que la mise en réseaux de nombreuses plate-forme boursières offrira des espaces de déploiement particulièrement intéressants. J'ai souhaité rendre accessible cette discipline peu connue et peu documentée dans notre pays, la France étant un des plus grands pays pourvoyeurs d'experts en finance quantitative grâce à la qualité des ses écoles et de ses formations.

BREF HISTORIQUE
Les origines (1860-1970)

Le trading algorithmique est récent si on le compare à l'existence tricentenaire de la bourse, et retracer son histoire implique plusieurs disciplines. Un des points de départ pourrait être fixé par la naissance du « pantélégraphe » de l'abbé Giovanni Caselli (On peut voir au Conservatoire national des arts et métiers de Paris deux exemplaires du « pantélégraphe ». A la réception, on utilisait une feuille de papier imbibée d'iodure de potassium pour faire apparaître la signature sous forme d'un dessin de couleur marron. Le traitement de l'image se faisait ligne par ligne. Ce traitement par ligne été employé pour la télévision et le fax). Exploité par la Société des télégraphes, de 1867 à 1870, il servait à expédier des ordres de Bourse et permettait d'authentifier la signature d'un donneur d'ordres. C'est cette accélération du traitement et de la diffusion de l'information qui va alors sans cesse augmenter, soit par la mise en relation d'acteurs éloignés soit par la rapidité d'exécution et de traitement. Les transactions se font ensuite par téléphone, par fax et en France par minitel.

Le trading automatisé (1970-1990)

A partir de cette période apparaît le trading télématique avec les premiers ordinateurs et les écrans de présentation des courbes. Dans un premier temps l' « automated trading » n'est pas un trading réalisé par des machines autonomes, mais la mise en place d'instruments diffusant automatiquement l'information pour les acheteurs et les vendeurs, ainsi que la dématérialisation du passage des ordres. La traduction exacte serait « informatisation des salles de marché». Cette informatisation a permis d'effectuer des calculs et des arbitrages systématiques sur les valeurs et les indices, méthode qui a pris le nom de « program trading ». En rendant le jeu de l'offre et de la demande beaucoup plus réactif, les systèmes d'informations ont accentué le crack boursier d'octobre 1987 aux USA. Les informations fournies à toutes les parties au même instant ont produit des mouvements de masse et des variations extrêmement amples des marchés (le 19 octobre 1987, Wall Street perd 22,6 %) [5].

Le trading algorithmique (1990-2010)

A partir de cette date les salles de marché ont commencé à s'informatiser massivement afin de proposer une cotation en continue et d'offrir une capacité de gestion des transactions toujours plus importantes. A partir de 1997, les ECN (Electronic Communication Network) sont conçus. Il s'agit de centres informatiques qui mettent en relation les professionnels et les brokers pour effectuer des transactions informatisées. Ces ordinateurs tournent 24h/24 et offrent une automatisation du marché. Tous les acteurs autorisés peuvent se connecter et effectuer des transactions sur le marché coté ou sur les marchés hors cote.

Les nouvelles possibilités de trading offertes par les ECN ont transformé le program trading en algorithmique trading. Les capacités des ordinateurs étant dorénavant suffisantes pour effectuer des calculs en temps réel, l'ensemble du marché s'est engouffré dans l'amélioration du trading assisté par des agents autonomes, les recherches s'effectuant dans les grandes institutions financières aussi bien que chez les éditeurs, les brokers ou les universités. L'idée est d'optimiser les traitements décisionnels ainsi que les coûts transactionnels, car les volumes de données croissent constamment et nécessitent une réactivité extrême et une analyse en continu. Les fonctions de trader évoluent donc vers plus de monitoring, les opérations les plus simples étant laissées aux algorithmes.

Parallèlement à ces développements tournés vers les besoins opérationnels des traders, depuis 2002 ce sont tenus plusieurs concours de trading automatique regroupant des équipes d'étudiants et de chercheurs autour de la plateforme de simulation Penn Lehmann Automatic Trading (PLAT) à l'université de Pennsylvanie (Kearns , M. ; Ortiz, L. (2003) - The Penn-Lehman Automated Trading Project. IEEE Intelligent Systems, 6, pp. 22-31, Nov/Dec).

Actuellement plusieurs groupes effectuent des recherches sur les algorithmes décisionnels automatisés : les chercheurs d'ITG Inc. autour de I. Domowitz, ensuite la communauté des chercheurs de l'Université du Texas à Austin avec M. Kearns, P. Stone, A. Sherstov, enfin au Royaume-Uni le Center for Financial Research de l'Université de Cambridge avec M. A. H. Dempster, R. G. Bates et V. Leemans.

BREF ETAT DE L'ART

La plupart des 31 éditeurs référencées par le journal Wall Street & Technology proposent des automates et des algorithmes. Tous offrent des algorithmes d'assistance aux traders tels que TWAP et VWAP qui permettent de réaliser de nombreuses transactions de faibles volumes avec une grande rapidité d'exécution, l'essentiel étant alors de maitriser l'optimisation des coûts de transaction et de vitesse d'exécution « temps réel ».

S'il y a de nombreux algorithmes et automates utilisés dans les salles de marché, les banques se refusent souvent à communiquer autour de ces technologies, ces informations étant considérées comme confidentielles. Seuls les éditeurs de logiciel, les cabinets de consultant (TABB Group) ou les laboratoires de recherches universitaires offrent des données sur l'état actuel des produits et des technologies.

Les algorithmes de base

Ces algorithmes fonctionnent tous sur le principe de la troncature des ordres volumineux en série de lots assimilables par le marché. Plusieurs avantages sont liés à cette technique : réduire les coûts de brokerage en faisant traiter les ordres par des machines aussi bien du côté achat que du côté vente au lieu de les passer par téléphone, construire des stratégies de vente aussi proche du marché que possible sur une certaine période afin de pouvoir vendre l'ensemble des lots, réduire les erreurs d'exécutions tout en épargnant de fastidieuses opérations aux traders.

L'utilisation de ce type d'algorithme fait plus particulièrement l'objet de séminaires et de formations car ces méthodes sont de plus en plus utilisées par les hedge funds et nécessitent une formation spécifique.

- VWAP (volume-weighted average price) : cette technique est utilisée principalement par les fonds de pension ou les fonds mutualistes pour vendre un gros volume en de nombreux petits ordres. Elle permet de réduire les coûts des brokerage par un pré-appariement des ordres d'achat/vente. Une transaction en VWAP se fera par exemple 40% le matin et 60% l'après midi, répartie selon le volume de transactions observées sur la place boursière.

- TWAP (time weighted average price) : cette technique est employée pour effectuer de nombreuses petites opérations selon une période donnée. Comme pour le VWAP il s'agit d'automatiser l'achat ou la vente d'un grand nombre d'actions par morceau. Un TWAP sur une journée se répartira 50% le matin et 50% l'après midi. Cette méthode permet d'identifier un prix de vente moyen.

- Arrival Price / Shortfall (IS, Implementation shortfall) : L'algorithme de prix d'arrivée permet de déterminer un prix fixé auquel on souhaite vendre ou acheter des actions. Le logiciel effectue toutes les transactions en tenant compte de l'impact sur le marché, de la liquidité, du volume et de la durée pour aboutir à ce prix moyen.

- Percent of Volume : Même principe de fractionnement des volumes cette fois ci en de manière proportionnelle au volume des transactions en cours ce qui permet de garantir une meilleure exécutions des ordres.

- TVOL (Target Volume) : Cet algorithme a les mêmes fonctionnalités que les précédents (VWAP, TWAP) mais effectue les transactions en fonction d'un volume d'achat ou de vente souhaité.

Fonctionnement

Ces algorithmes étant paramétrés ils doivent être accessibles aux programmeurs qui vont leur donner des instructions de travail. Les principales données employées ne varient pas de celles déjà utilisées par les traders, ces algorithmes étant basés sur des stratégies de trading conçues et expérimentées en salle de marché entre les Quants et les traders.

Les automates peuvent fonctionner sur différentes OS, Windows/Mac/Unix et sont programmés avec différents langages Java, C#, et principalement C++. Ils emploient aussi des systèmes de cryptage et de certificats digitaux qui servent à authentifier et sécuriser les transactions. Les machines qui doivent supporter ces algorithmes sont particulièrement puissantes, car dans certains cas elles effectuent des traitements sur plusieurs sources en temps réel en tenant compte des données historiques afin de calibrer les transactions (volatilité implicite pour le trading option par exemple).

L'idée de départ est d'obtenir un agent autonome dans ces décisions de trading afin d'effectuer des transactions de manière extrêmement réactive. Pour cela il faut sélectionner les données que l'automate devra prendre en compte. La première donnée est le positionnement dans le temps. Quelles sont les périodes optimums à prendre en compte pour construire un jeu de données fiable ; à quel moment initier ou clore une transaction. La seconde donnée est quelles types de transactions effectuer et sur quels marchés et en se basant sur quelles données : écart-type (volatilité), microstructure de marché, pas de la cotation (spread). Enfin la troisième donnée concerne les contraintes à mettre en place pour éviter des pertes importantes.

Pour optimiser les coûts de transactions les Brokers ont proposés des accès direct au marché (DMA) spécialement dédiés aux transactions électroniques. Le trader passe de moins en moins d'ordres par téléphone, et se concentre sur des transactions spécifiques. Il existe plusieurs fournisseurs d'algorithmes :

- les Brokers disposant de plate-forme DMA offrent en général des algorithmes à l'usage de leurs clients
- les éditeurs qui peuvent fournir des algorithmes personnalisés
- le développement en interne par les plus grandes banques.

La puissance des ordinateurs permet actuellement d'implémenter de plus en plus efficacement des algorithmes comportementaux et décisionnels. Basés sur les concepts d'intelligence artificielle et de la théorie des jeux, ces algorithmes nécessitent d'être pilotés et paramétrés par des quant et des traders toujours plus compétent en informatique et en mathématique.

Certaines personnes pensent que ces machines assureront 40% des échanges de capitaux sur les marchés des equities américains d'ici 2008-2010 (Der Hovanesian, Mara. (2006) - Cracking The Street's New Math. NY financial).

CONCLUSION

Quand les algorithmes de trading sont utilisés pour découper les ordres en petites transactions le premier effet est de produire une surcharge du nombre d'ordres passés, obligeant les bourses à mettre en œuvre des plateformes toujours plus puissantes pour éviter les erreurs de cotations et de transactions. A cette fin, des éditeurs comme IBM proposent des technologies spécifiques pour la finance avec des serveurs 4 cœurs spécialement adaptés pour la gestion des flux.

Un autre effet secondaire est que ces technologies réclament des ressources humaines très spécialisées en finance quantitative et informatique. Depuis 2005 les Quant sont donc devenus particulièrement sollicités par les hedge funds et les grandes banques d'affaires afin d'effectuer des calculs sur les tendances à partir des données du marché ou de développer des calculs spécifiques. La demande se fait aussi sentir du coté des éditeurs et des prestataires de services informatiques dont les clients réclament des plateformes opérationnelles et l'intégration d'algorithmes toujours plus affinés. La vente d'algorithmes par les sociétés de courtages ou des éditeurs doit aussi faire face aux limites des clients qui ne sont pas toujours au fait des dernières applications de la recherche en finance quantitative et pourraient montrer des difficultés à utiliser des services trop complexes.

BIBLIOGRAPHIE

1. Azoff, E. *Neural Network Time Series Forecasting of Financial Markets*, Wiley & Sons, New York, NY: 1994.

2. Austin, M. P. ; Bates, R. G. ; Dempster, M. A. H. ; Leemans, V. ; Williams, S. N. Adaptive systems for foreign exchange trading. *Quantitative Finance*, 4(4), 2004, 37–45.

3. Bates, R. G. ; Dempster, M. A. H. ; Romahi, Y. S. Evolutionary reinforcement learning in FX order book and order flow analysis. *Proceedings of the 2003 IEEE International Conference on Computational Intelligence for Financial Engineering*, Hong Kong, March 2003, 355–362.

4. Bollerslev, Tim ; Domowitz, Ian ; Wang, Jianxin. Order flow and the bid-ask spread: An empirical probability model of screen -based trading. *Journal of Economic Dynamics and Control*, Volume 21, Issues 8-9, 29 June 1997, Pages 1471-1491.

5. Bouchard, J. ; Mezard, M. ; Potters, M. Statistical Properties of Stock OrderBooks: empirical Results andModels. *Quantitative Finance*, vol. 2, no. 4, Aug. 2002.

6. Brailsford, Timothy J. ; Frino, Alex ; Hodgson, Allan ; West, Andrew. Stock market automation and the transmission of information between spot and futures markets. *Journal of Multinational Financial Management*, Volume 9, Issues 3-4, November 1999, Pages 247-264.

7. Brock, W. ; Lakonishok, J. ; LeBaron, B. Simple technical trading rules and the stochastic properties of stock returns. *J. Finance*, vol. 47, no. 5, 1992, pp. 1731-1763.

8. Chiu, Kai-Chun ; Xu, Lei. Arbitrage pricing theory-based Gaussian temporal factor analysis for adaptive portfolio management. *Decision Support Systems*, Volume 37, Issue 4, September 2004, Pages 485-500.

9. Coppejans, Marc ; Domowitz, Ian. Pricing behavior in an off-hours computerized market. *Journal of Empirical Finance*, Volume 6, Issue 5, December 1999, Pages

583-607.

10. Dempster, M.A.H. ; Leemans, V. An automated FX trading system using adaptive reinforcement learning. *Expert Systems with Applications*, Volume 30, Issue 3, April 2006, Pages 543-552.

11. Dempster, M. A. H. ; Romahi, Y. S. Intraday FX trading: An evolutionary reinforcement learning approach. *Intelligent Data Engineering and Automated Learning*, 2002, Lecture Notes in Computer Science. Berlin: Springer pp. 347–358.

12. Dempster, M. A. H. ; Payne, T. W. ; Romahi, Y. S. ; Thompson, G. W. P. Computational learning techniques for intraday FX trading using popular technical indicators. *Special issue on Computational Finance, IEEE Transactions on Neural Networks*, 12, 2001, 744–754.

13. Dempster, M. A. H., & Jones, C. M. A real-time adaptive trading system using genetic programming. *Quantitative Finance*, 1, 2001, 397–413.

14. Domowitz, I. ; Yegerman, H. Measuring and Interpreting the Performance of Broker Algorithms. ITG Inc. research report, August 2005.

15. Domowitz, I., Yegerman, H. The Cost of Algorithmic Trading: A First Look at Comparative Performance. *Brian Bruce, ed., Algorithmic Trading: Precision, Control, Execution*, New York: Institutional Investor, 2005.

16. Domowitz, Ian ; Hansch, Oliver ; Wang, Xiaoxin. Liquidity commonality and return co-movement. *Journal of Financial Markets*, Volume 8, Issue 4, November 2005, Pages 351-376.

17. Domowitz, Ian ; Glen, Jack; Madhavan, Ananth. Liquidity, Volatility, and Equity Trading Costs Across Countries and Over Time. *International Finance,* 4, 221-256, 2001.

18. Domowitz, Ian ; El-Gamal, Mahmoud A. A consistent nonparametric test of ergodicity for time series with applications. *Journal of Econometrics*, Volume 102, Issue 2, June 2001, Pages 365-398.

19. Domowitz, Ian. Electronic derivatives exchanges: Implicit mergers, network externalities, and standardization. *The Quarterly Review of Economics and Finance*, Volume 35, Issue 2, Summer 1995, Pages 163-175.

20. Domowitz, Ian ; Wang, Jianxin. Auctions as algorithms : Computerized trade execution and price discovery. *Journal of Economic Dynamics and Control*, Volume 18, Issue 1, January 1994, Pages 29-60.

21. Domowitz, Ian. A taxonomy of automated trade execution systems. *Journal of International Money and Finance*, Volume 12, Issue 6, December 1993, Pages 607-631.

22. Domowitz, Ian. The mechanics of automated trade execution systems. *Journal of Financial Intermediation*, Volume 1, Issue 2, June 1990, Pages 167-194.

23. Donaldson, R. ; Kamstra, M. Forecast combining with neural Network. *Journal of Forecating*, Vol. 15, No. 1, pp. 49-61, 1996.

24. Feng, Y. ; Yu, R. ; Stone, P. Two Stock-Trading Agents: Market Making and Technical Analysis. *Proc of AMEC V Workshop*, Springer LNAI, pp. 18-36, 2004.

25. Ferris, Stephen P. ; McInish, Thomas H. ; Wood, Robert A. Automated trade execution and trading activity: The case of the Vancouver stock exchange. *Journal of International Financial Markets, Institutions and Money*, Volume 7, Issue 1, April 1997, Pages 61-72.

26. Freedman R. et. Al. Artificial intelligence in Capital Markets, Chicago, IL : Probus Pud., 1995.

27. Gately, E. *Neural Networks for Financial Forecasting*, Wiley & Sons, New York, NY : 1996.

28. Giraud, Jean-Renĕ. Best Execution for Buy-Side Firms: A Challenging Issue, A Promising Debate, A Regulatory Challenge. *consulting report.* Edhec-Risk Advisory, June 2004.

29. Gold, C. FX trading via recurrent reinforcement learning. *Proceedings of the IEEE International Conference on Computational Intelligence in Financial Engineering*, 2003, op.cit., 363–371.

30. Grossman, Randy. The Search for the Ultimate Trade: Market Players in Algorithmic Trading. *consulting report.* Financial Insights, February, 2005.

31. Harrison, J.M. ; Kreps, D. Martingales and arbitrage in multiperiod securities markets. *J. Econ. Theory*, vol. 20, 1979, pp. 381-408.

32. Hogan, Steve ; Jarrow, Robert ; Teo, Melvyn ; Warachka, Mitch. Testing market efficiency using statistical arbitrage with applications to momentum and value strategies. *Journal of Financial Economics* 73, 2004, 525–565.

33. Ingber, L. ; Mondescu, R.P. Optimization of trading physics models of markets. *IEEE Trans. on Neural Networks*, vol. 12, no. 4, 2001, pp. 776-790, http://www.ingber.com/markets01 optim trading.pdf .

34. Ingber, L. High-resolution path-integral development of financial Options. *Physica A*, vol. 283, no. 3/4, 2000, pp. 529-558, http://www.ingber.com/markets00 highres.ps.gz

35. Ingber, L. Statistical mechanics of nonlinear nonequilibrium financial markets: applications to optimized trading. *Math. Computer Modelling*, vol. 23, no. 7, 1996, pp. 101-121, http://www.ingber.com/markets96 trading.ps.gz .

36. Ingber, L. ; Rosen, B. Genetic algorithms and very fast simulated reannealing: a comparison. *Oper. Res. Management Sci.*, vol. 33, no. 5, 1993, p. 523.

37. Ingber, L. ; Wehner, M.F. ; Jabbour, G.M. ; Barnhill, T.M. Application of statistical mechanics methodology to term-structure bond-pricing models. *Math. Comput. Modelling*, vol. 15, no. 11, 1991, pp. 77-98, http://www.ingber.com

/markets91 interest.ps.gz .

38. Kakade, S. ; Kearns, M. ; Mansour, Y. ; Ortiz, L. Competitive Algorithms for VWAP and Limit Order Trading. *Proc. ACM Conf. Electronic Commerce*, 2004.

39. Kaufman, P.J. *Trading Systems and Methods*, 3rd ed., John Wiley & Sons, New York, NY : 1998.

40. Kearns, M. ; Ortiz, L. The Penn-Lehman Automated, Trading Project. *IEEE Intelligent Systems*, No. 6, pp. 22-31, Nov/Dec 2003.

41. Kingdon, J. Intelligent Systems and financial forecasting. New York, NY : Springer, 1997.

42. Leemans, V. Real time trading systems. MPhil dissertation, Centre for Financial Research, Judge Institute of Management. Cambridge: University of Cambridge, 2003.

43. Lucchetti, Aaron ; Lahart, Justin. Your Portfolio on Autopilot, *The Wall Street Journal*. Sep 30, 2006

44. Madhavan, Ananth ; Porter, David ; Weaver, Daniel. Should securities markets be transparent? *Journal of Financial Markets*, Volume 8, Issue 3, August 2005, Pages 265-287.

45. Mandelbrot, B.B. *Fractals and Scaling in Finance*, Springer- Verlag, New York, NY : 1997.

46. Masters, Timothy. Just what are we optimizing, anyway? *International Journal of Forecasting*, Volume 14, Issue 2, 1 June 1998, Pages 277-290.

47. Moody, J. ; Saffell, M. Learning to trade via direct reinforcement. *IEEE Transactions on Neural Networks*, 12(4), 2001, 875–889.

48. Moody, J. ; Saffell, M. Minimising downside risk via stochastic dynamic programming. In Y. S. Abu-Mostafa, B. LeBaron, A. W. Lo, & A. S. Weigend (Eds.), *Sixth International conference computational finance,* 1999, pp. 403–415. Cambridge, MA: MIT Press, 1999.

49. Naidu, G. N. ; Rozeff, Michael S. Volume, volatility, liquidity and efficiency of the Singapore Stock Exchange before and after automation. *Pacific-Basin Finance Journal*, Volume 2, Issue 1, March 1994, Pages 23-42.

50. Ogryczak, W., & Ruszczyn´ski, A. Dual stochastic dominance and related mean-risk models. SIAM. *Journal on Optimization*, 13(1), 2002, 60–78.

51. Ogryczak, W., & Ruszczyn´ski, A. From stochastic dominance to meanrisk models: Semideviations as risk measures. *European Journal of Operations* Research, 116, 1999, 33–50.

52. Perold, A. The Implementation Shortfall: Paper versus Reality. *The Journal of*

*Portfolio Management*, vol. 14, no. 3, Spring 1988, p. 49.

53. QSG. The implementation Costs of Algorithmic Trading. *consulting report.* Quantitative Services, Group LLC., December 2004.

54. Ruszczyn´ski, A. ; Vanderbei, R. J. Frontiers of stochastically nondominated portfolios. *Econometrica*, 71(4), 2003, 1287–1297.

55. Sherstov, A. ; Stone, P. Three Automated Stock Trading Agents: A Comparative Study. *Proc. of the AMEC VI Workshop*, New York, USA, July 2004.

56. Silaghi, Gheorghe Cosmin ; Bolyai, Babes. An Agent Strategy for Automated Stock Market Trading Combining Price and Order Book Information. University Faculty of Economic Sciences, Cluj-Napoca, Romania

57. Skabar, A. ; Cloete, I. Discovery of Financial Trading Rules. *Proc. Artificial Intelligence and Applications*, pp. 121-125, 2001.

58. Stone, P. ; Littman, M. ; Singh, S. ; Kearns M. ATTac-2000: An adaptive autonomous bidding agent. Journal of Artificial Intelligence Research, no. 15, June 2001, pp. 189-206.

59. Tabb, Larry. Institutional Equity Trading in America: A Buy-Side Perspective. *consulting report*, The Tabb Group, April 2004.

60. Venturi S. Evolutionary algorithms for currency trading. PhD thesis, Centre for Financial Research, Judge Institute of Management, University of Cambridge, 2003.

61. Wofsey, M. Technology: shortcut tests validity of complicated Formulas. *The Wall Street Journal*, vol. 222, no. 60, 1993, p.B1.

62. Wurman, P.; Wellman, M. ; Walsh W. The Michigan Internet AuctionBot: a configurable server for human and software agents. *Second international conference on autonomous agents*, Mineapolis, pp 301-308.

63. Xufre Casqueiro, Patrícia ; Rodrigues, António J.L. Neuro-dynamic trading methods. *European Journal of Operational Research*, Volume 175, Issue 3, 16 December 2006, Pages 1400-1412.

64. Yang, J., and M. Borkovec. Algorithmic Trading: Opportunities and Challenges. *Financial Engineering News*, No. 46, November/December 2005, pp.14-15.

65. Ygge, F. Market-oriented programming and its application to power load management. Ph.D. Thesis, Lund University, 1998.

66. Yu, Ronggang ; Stone, Peter. Performance Analysis of a Counterintuitive Automated StockTrading Agent. Department of Computer Sciences, The University of Texas at Austin, www.cs.utexas.edu/~ryu